\documentclass[aps,prc,preprint,superscriptaddress]{revtex4}
\usepackage{graphicx}
\usepackage{dcolumn}
\usepackage{bm}
\begin{document}

\setcounter{page}{1} \pagestyle{myheadings} \oddsidemargin
4mm\evensidemargin -4mm

\title{A Simple Method for Rise-Time Discrimination of Slow Pulses from
Charge-Sensitive Preamplifiers}

\author{Jan T\~oke}
\affiliation{Department of Chemistry, University of Rochester,
Rochester, New York 14627}
\author{Michael J. Quinlan}
\affiliation{Department of Chemistry, University of Rochester,
Rochester, New York 14627}
\author{Wojtek Gawlikowicz}
\affiliation{Department of Chemistry, University of Rochester,
Rochester, New York 14627}
\author{W. Udo Schr\"oder}
\affiliation{Department of Chemistry, University of Rochester,
Rochester, New York 14627} \affiliation{Department of Physics and
Astronomy, University of Rochester, Rochester, New York 14627}
\begin{abstract}

Performance of a simple method of particle identification via pulse
rise time discrimination is demonstrated for slow pulses from
charge-sensitive preamplifiers with rise times ranging from 10 ns to
500 ns. The method is based on a comparison of the amplitudes of two
pulses, derived from each raw preamplifier pulse with two amplifiers
with largely differing shaping times, using a fast peak-sensing ADC.
For the injected charges corresponding to energy deposits in silicon
detectors of a few tens of MeV, a rise time resolution of the order
of 1 ns can be achieved. The identification method is applicable in
particle experiments involving large-area silicon detectors, but is
easily adaptable to other detectors with a response corresponding to
significantly different pulse rise times for different particle
species.

\end{abstract}

\maketitle

\section{Introduction}

Pulse-shape discrimination is a well known technique often used to
identify particles according to their atomic and mass numbers. The
method is based on the characteristic time-dependence of the
electrical or optical response they generate in the detectors. There
are different physical effects responsible for the sensitivity of
different detectors on the species of impinging particles.
Accordingly, many different pulse-shape discrimination methods have
been devised.In particular, the sensitivity of semiconductor
detectors to different particles is due to the variation of the
ionization density along the particle track in the detector material
and the resulting temporal variation of the charge collection rate
in the applied electric field.

The present study is inspired by the actual demand of measuring rise
times of pulses produced by charged particles in large-area silicon
detectors now used in multidetector arrays such as the CHIMERA
telescope array \cite{chimera}. The study focuses on an idea of
measuring rise times of slow pulses from charge-sensitive
preamplifiers by shaping (filtering) these pulses with two different
timing constants, one of which is much longer than the relevant
maximum rise time, while the other is of the order of the shortest
rise time expected in the particular application. With proper
shaping, the amplitude of the ``slow'' pulse (representing the total
charge injected into the preamplifier) will be virtually independent
of the rise time of the raw preamplifier pulse, while the amplitude
of the ``fast'' pulse (representing the fraction of charge injected
on a short timescale) will show a distinct dependence on the rise
time of the preamplifier pulse. Subsequently, the amplitudes of both
shaped pulses can be digitized by fast peak-sensing Analog to
Digital Converters (ADCs), such as, e.g., the Phillips 7164
\cite{phillips} or Silena \cite{silena} modules. Alternatively,
depending on the available data acquisition, both shaped pulses can
be stretched and their amplitudes be measured using
current-integrating ADCs (QDC). Ultimately, the rise time is
determined, event-by-event, from the ratio of ``fast'' to ``slow''
pulse amplitudes.

\section{The Test Setup}

\begin{figure}
\includegraphics[width=15cm]{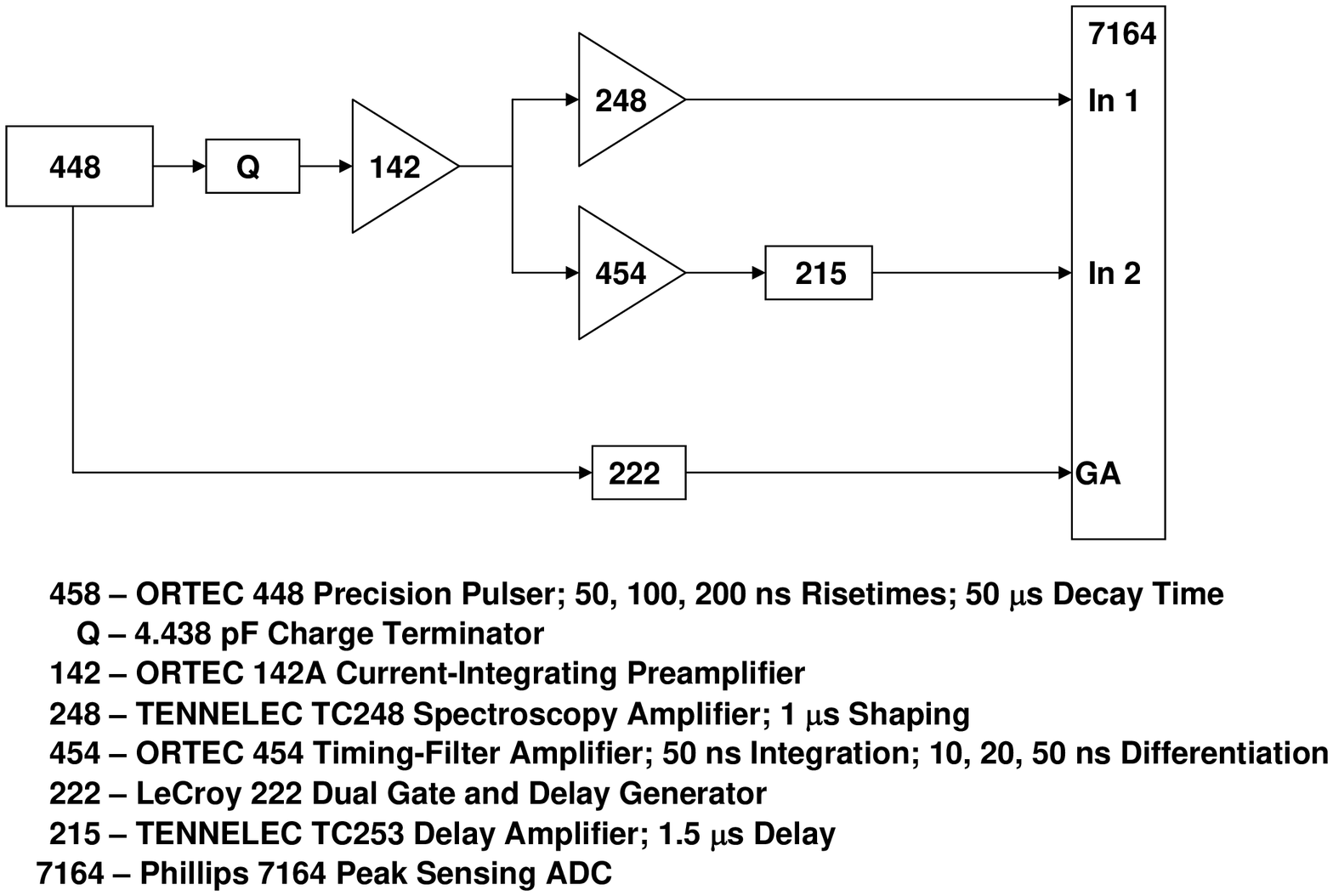}
\caption{Block diagram of the electronic setup used to determine the
performance of the pulse-shape discrimination
method.}\label{fig:setup}
\end{figure}

\begin{figure}
\includegraphics[width=10cm]{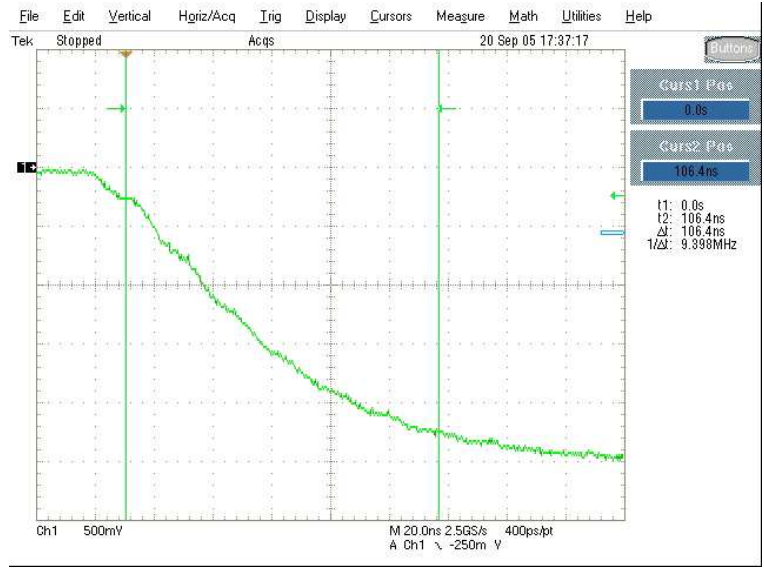}
\caption{Appearance of the leading edge of a typical pulse from
the charge-sensitive preamplifier with nominally 100 ns rise
time.}\label{fig:pulse}
\end{figure}

A block diagram of the test setup is shown in Fig.~\ref{fig:setup}.
As depicted in this diagram, calibrated charge was injected into an
ORTEC 142A charge-sensitive preamplifier, with three different
rates, resulting in preamplifier output signals that had nominal
rise times of 50, 100, and 200 ns, respectively. Charge pulses were
obtained by charge-terminating pulses from a precision ORTEC 448
pulse generator. Different pulse rise times were employed as defined
by the rise time settings of the pulse generator, while the decay
time was fixed at 50 $\mu$s. A typical pulse with a nominal (set on
the pulse generator module) rise time of 100 ns is illustrated in
Fig.~\ref{fig:pulse}. It is seen to have an actual rise time close
to nominal.

The output signal from the preamplifier was routed into two
spectroscopic channels characterized by different shaping times. The
``slow'' channel had a shaping time fixed at 1 $\mu$s, while for the
``fast'' channel differentiation times of 10, 20, and 50 ns were
explored. The integration time of the ORTEC 454 timing filter
amplifier (TFA) used in the ``fast'' channel was fixed at 50 ns.

\begin{figure}
\includegraphics[width=10cm]{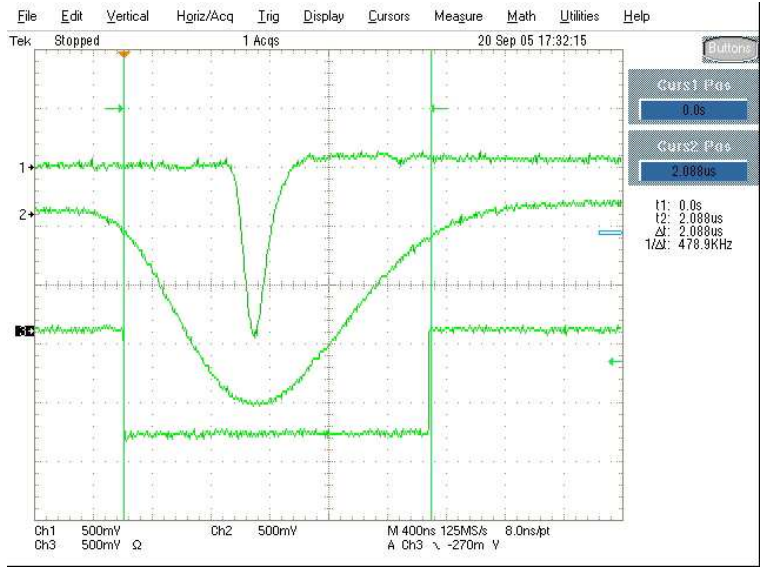}
\caption{The timing diagram of the ``fast'' and ``slow''
spectroscopic pulses and of the ADC gating
pulse.}\label{fig:timing}
\end{figure}

To be able to use a common ADC gate for both, fast and slow pulses,
a TC215 delay amplifier was used in the fast channel to delay the
output pulse from the ORTEC 454 TFA. The resulting timing of fast
and slow pulses at the input of the peak sensing Phillips 7164 ADC
is illustrated in FIG.~\ref{fig:timing}. The present setup requires
a peak sensing ADC capable of digitizing pulses with 50-ns short
rise times, hence the choice of the Phillips 7164 ADC module.

\section{Results and Analysis}

The measurements were performed for series of pulses with rise times
of 50 ns, 100 ns, and 200 ns and for shaping times for the ``fast''
response of 10 ns, 20 ns, and 50 ns. To assess the effect of
inherent fluctuations in the amount of charge ($Q$)injected into the
preamplifier, the latter charge was varied in the range from 0.222
pC to 3.55 pC, corresponding to energy deposits in a silicon
detector in the range from 5 MeV to 80 MeV.

\begin{figure}
\includegraphics[width=10cm]{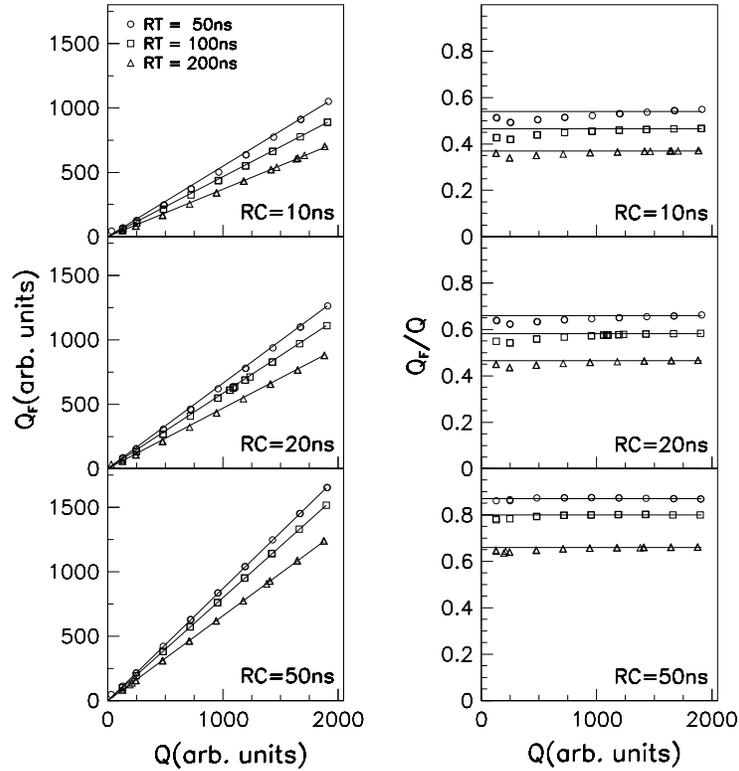}
\caption{The dependence of the average amplitude ($Q_F$)of the
``fast'' pulse on the average height ($Q$) of the ``slow'' pulse for
different amounts of injected charge, different rise times of the
raw preamplifier pulse, and different shaping times, as indicated in
the panels. The right column displays the corresponding ratios
$Q_f/Q$.}\label{fig:fastvsslow}
\end{figure}

\begin{figure}
\includegraphics[width=10cm]{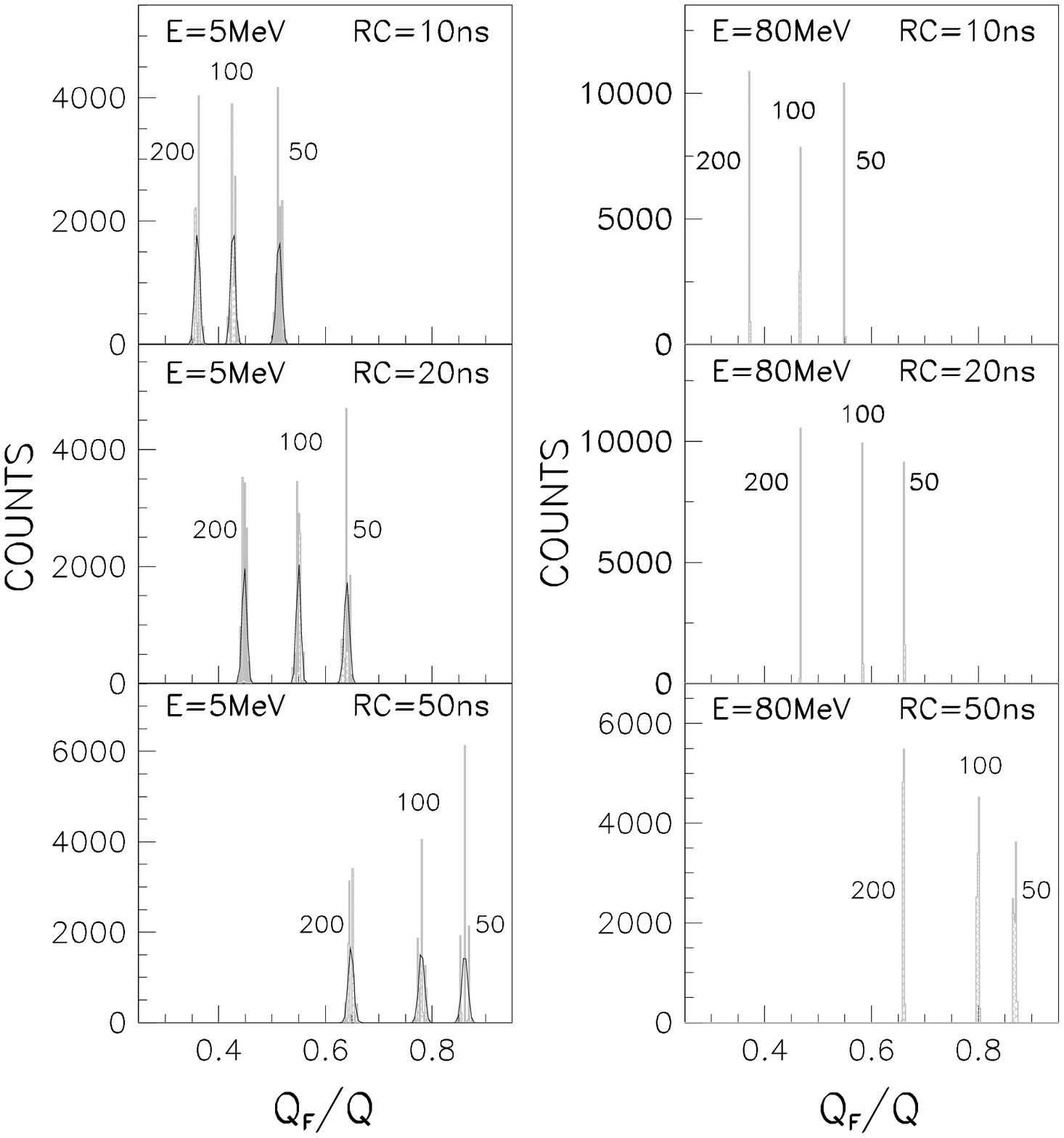}
\caption{Spectra of the amplitude ratios ($Q_F/Q$) of ``fast''
($Q_F$) and ``slow'' ($Q$) pulses, measured with various $RC$
shaping times and two representative injected charges. The three
peaks seen in every panel correspond to the different rise times
indicated by labels.}\label{fig:peaks}
\end{figure}

\begin{figure}
\includegraphics[width=10cm]{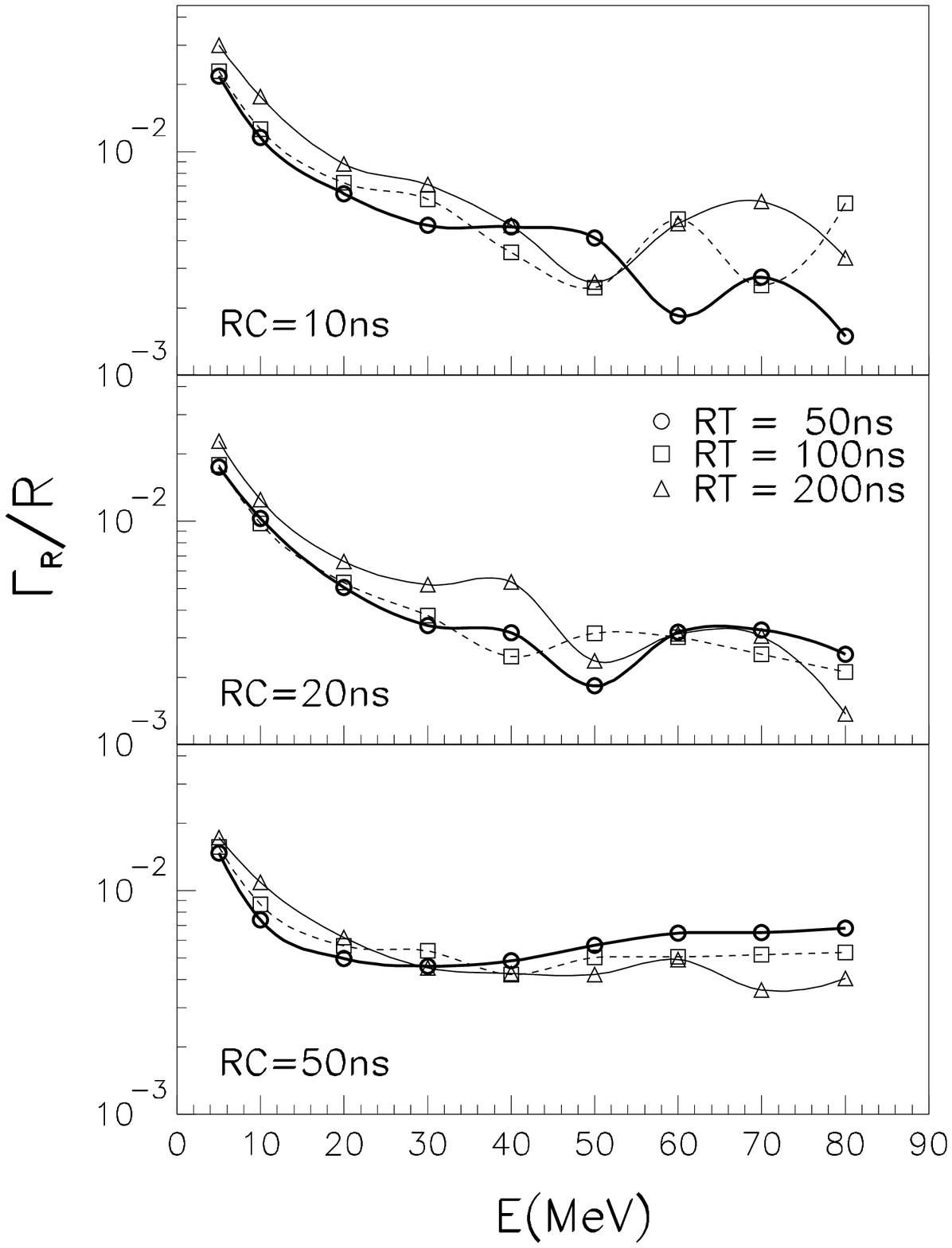}
\caption{Relative resolution obtained for amplitude ratios of
``fast'' and ``slow'' pulses as functions of injected charge for
different $RC$ shaping times and rise times ($RT$, as indicated by
labels. The injected charge is represented by the corresponding
energy deposit $E$ in a silicon detector, plotted on the
abscissa.}\label{fig:relgamma}
\end{figure}

\begin{figure}
\includegraphics[width=10cm]{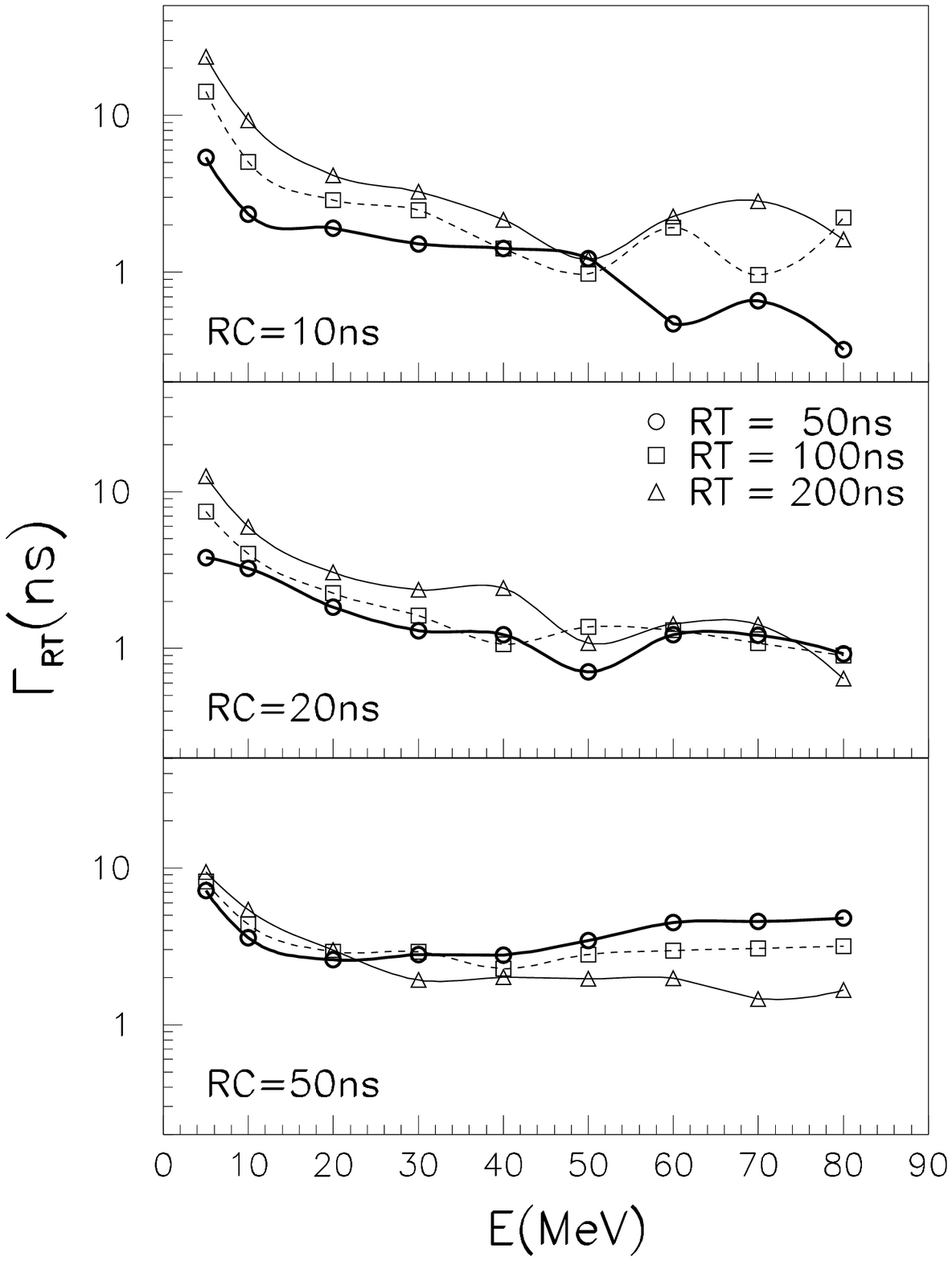}
\caption{Absolute resolution (FWHM) in rise time $RT$, as extracted
from data for different amounts of injected charge and different
shaping and rise times, as indicated by labels.}\label{fig:absgamma}
\end{figure}

Results of the measurements are illustrated in
Figs.~\ref{fig:fastvsslow}, \ref{fig:peaks}, \ref{fig:relgamma}, and
\ref{fig:absgamma}. Fig.~\ref{fig:fastvsslow} illustrates the
dependence of the amplitude ($Q_F$) of the ``fast'' pulse on the
injected charge ($Q$), as measured for different shaping and rise
times denoted by time constants $RC$ and $RT$, respectively. As seen
in this figure, shorter $RC$ shaping times provide for a better
separation of faster pulses, a trend that agrees qualitatively with
what is expected based on the principles of filtering.

A more quantitative measure of the pulse-shape discrimination
performance of the method over a range of injected charges is seen
in Fig.~\ref{fig:peaks}, where spectra of the ratios of the
amplitudes ($Q_F$) of ``fast'' pulses to the amplitudes ($Q$) of the
corresponding ``slow'' pulses are displayed for various $RC$ shaping
times. As expected, the resolution of the method deteriorates with
decreasing amount of injected charge. Again, short shaping times are
seen to provide for a larger spread between pulses of short rise
times, at the expense of a loss in the amplitude ($Q_F$) of ``fast''
pulses.

The relative widths $\Gamma_R/R$ of the peaks seen in
Fig.~\ref{fig:peaks} are plotted in Fig.~\ref{fig:relgamma} {\it
vs.} injected charge for different shaping and rise times. The
charge is expressed in units of MeV of equivalent energy deposit for
silicon. As seen in this figure, the relative resolution is almost
equally good for 10-ns and 20-ns shaping times. The resolution is
better than 1\% over most of the range of injected charges, except
for the smallest one.

Fig.~\ref{fig:absgamma} illustrates the absolute resolution
$\Gamma_{RT}$ in rise time in units of nanoseconds, achieved for
different amounts of injected charge ($E$) and for different shaping
and rise times. As seen in this figure, for the range of rise times
explored, the 10-ns and 20-ns shaping times offer an almost equally
good resolution of a few ns, over much of the range in charges
explored. The weak but steady deterioration of the resolution with
decreasing amount of charge injected, comes from the relative
increase in statistical fluctuation in charge. It is encouraging
that, except for the lowest charges injected, the resolution is of
the order of a few ns, and is approximately 1 ns for the range of
energies involved in typical experiments.

\section{Conclusions}

A simple and economical method has been demonstrated for rise time
discrimination of pulses from charge sensitive preamplifiers. The
method is based on measurements of the amplitudes of pulses derived
from the raw preamplifier pulses by shaping with different time
constants. It is shown to provide for a very good resolution in rise
times. The method requires no fine tuning of the electronics
circuitry involved and should, therefore, be well suited for
multi-detector systems. In the present study, a timing filter
amplifier was used to produce the ``fast'' signal. In fact, a most
primitive RC differentiator should perform sufficiently well,
provided its output amplitude is matched to the range of the
available ADC. The present design processes pulses from
charge-sensitive preamplifiers used with solid-state detectors, but
the method can easily be adapted to other systems such as
scintillation detectors providing pulses with significantly
different rise times.

\begin{acknowledgments}
This work was supported by the U.S. Department of Energy grant No.
DE-FG02-88ER40414.
\end{acknowledgments}

\end{document}